\documentclass{pasj00}

\usepackage{natbib}
\newcommand{\pasa}{PASA}

\begin{document}
\SetRunningHead{Doi et al.}{Multifrequency VLBI observations of the BAL quasar J1020+4320}
\Received{2012/11/16}
\Accepted{2012/12/31}

\title{Multifrequency VLBI Observations of the Broad Absorption Line Quasar J1020+4320: Recently Restarted Jet Activity?}

\author{
Akihiro \textsc{Doi}\altaffilmark{1,2}, 
Yasuhiro \textsc{Murata}\altaffilmark{1,2}, 
Nanako \textsc{Mochizuki}\altaffilmark{1}, 
Hiroshi \textsc{Takeuchi}\altaffilmark{1}, \\
Keiichi \textsc{Asada}\altaffilmark{3}, 
Takayuki J. \textsc{Hayashi}\altaffilmark{4,5},  
Hiroshi \textsc{Nagai}\altaffilmark{6}, 
Katsunori~M. \textsc{Shibata}\altaffilmark{4}, \\
Tomoaki \textsc{Oyama}\altaffilmark{4}, 
Takaaki \textsc{Jike}\altaffilmark{4}, 
Kenta \textsc{Fujisawa}\altaffilmark{7,8}, 
Koichiro \textsc{Sugiyama}\altaffilmark{9}, \\
Hideo \textsc{Ogawa}\altaffilmark{10}, 
Kimihiro \textsc{Kimura}\altaffilmark{10}, 
Mareki \textsc{Honma}\altaffilmark{4,11}, 
Hideyuki \textsc{Kobayashi}\altaffilmark{4,5}, \\
and 
Shoko \textsc{Koyama}\altaffilmark{4,5}  
}

\altaffiltext{1}{The Institute of Space and Astronautical Science, Japan Aerospace Exploration Agency,\\ 3-1-1 Yoshinodai, Chuou-ku, Sagamihara, Kanagawa 229-8510}
\altaffiltext{2}{Department of Space and Astronautical Science, The Graduate University for Advanced Studies,\\ 3-1-1 Yoshinodai, Chuou-ku, Sagamihara, Kanagawa 229-8510}
\altaffiltext{3}{Academia Sinica Institute of Astronomy and Astrophysics,\\ P.O. Box 23-141, Taipei 10617, Taiwan}
\altaffiltext{4}{Mizusawa VLBI Observatory, National Astronomical Observatory of Japan,\\ 2-21-1 Osawa, Mitaka, Tokyo 181-8588}
\altaffiltext{5}{Department of Astronomy, University of Tokyo, 7-3-1 Hongo, Bunkyo-ku, Tokyo 113-0033}
\altaffiltext{6}{National Astronomical Observatory of Japan, 2-21-1 Osawa, Mitaka, Tokyo 181-8588}
\altaffiltext{7}{Department of Physics, Faculty of Science, Yamaguchi University,\\ Yoshida 1677-1, Yamaguchi, Yamaguchi 753-8512}
\altaffiltext{8}{The Research Institute for Time Studies, Yamaguchi University,\\ Yoshida 1677-1, Yamaguchi, Yamaguchi 753-8511}
\altaffiltext{9}{Graduate school of Science and Engineering, Yamaguchi University,\\ 1677-1 Yoshida, Yamaguchi, Yamaguchi 753-8512}
\altaffiltext{10}{Department of Physical Science, Osaka Prefecture University,\\ 1-1 Gakuen-cho, Naka-ku, Sakai 599-8531}
\altaffiltext{11}{Department of Astronomical Science, Graduate University for Advanced Studies,\\ 2-21-1 Osawa, Mitaka, Tokyo 181-8588}

\KeyWords{galaxies: active --- galaxies: jets --- quasars: absorption lines --- radio continuum: galaxies --- galaxies: quasars: individual (SDSS J102027.20+432056.2)}

\maketitle

\begin{abstract}
This paper reports very-long-baseline interferometry observations of the radio-loud broad absorption line~(BAL) quasar J1020+4320 at 1.7, 2.3, 6.7, and 8.4~GHz using the Japanese VLBI network~(JVN) and European VLBI network~(EVN).  
The radio morphology is compact with a size of $\sim10$~pc.  The convex radio spectrum is stable over the last decade; an observed peak frequency of 3.2~GHz is equivalent to 9.5~GHz in the rest frame, suggesting an age of the order of $\sim100$~years as a radio source, according to an observed correlation between the linear size and peak frequency of compact steep spectrum~(CSS) and giga-hertz peaked spectrum~(GPS) radio sources.  
A low-frequency radio excess suggests relic of past jet activity.    
J1020+4320 may be one of the quasars with recurrent and short-lived jet activity during a BAL-outflowing phase.    
\end{abstract}

\section{INTRODUCTION}
Broad absorption line (BAL) quasars are identified in rest-frame ultra violet spectra by broad absorption troughs displaced blueward from the corresponding emission lines, such as C$_\mathrm{IV}$ and Mg$_\mathrm{II}$ \citep{Weymann:1991}.  
The blueward displacements, sometimes up to $\sim0.2c$, are attributed to intervening outflows from active galactic nuclei~(AGNs) in our line of sight.  
The intrinsic percentage of quasars with BALs is $\sim$20\% \citep{Hewett:2003,Knigge:2008}.   
Two plausible explanations, an orientation scheme and an evolutionary scheme, were proposed and have been long debated.      
The orientation scheme proposes that the outflowing BAL wind is preferentially equatorial \citep{Murray:1995,Proga:2000} and can be observed as absorption troughs only when the accretion disk is viewed nearly edge-on.    
The evolutionary scheme proposes that BAL outflows are associated with a relatively short-lived (possibly episodic) evolutionary phase \citep{Gregg:2000,Gregg:2006}.

Radio observations offer several inclination indicators and age estimators that can be used to test these two controversial schemes for BAL quasars \citep[][and references as follows]{Stocke:1992,Richards:2001,Menou:2001,Brotherton:2006,DiPompeo:2012}.  
Several radio sources in BAL quasars exhibit rapid radio variability that indicates unusually high brightness temperatures, which requires Doppler beaming of nearly pole-on viewed jets, i.e., accretion disks with small inclinations \citep{Zhou:2006a,Ghosh:2007}.  
A wide range of spectral indices, including flat spectra and steep spectra, is consistent with the wide range of orientations \citep{Becker:2000,Montenegro-Montes:2008,Fine:2011}.  
Although there is weak evidence that the spectral indices of BAL quasars are steeper than those of non-BAL quasars, which mildly favors edge-on orientations \citep{DiPompeo:2011,Bruni:2012}, a single edge-on geometry cannot describe all BAL quasars.    
\citet{Becker:2000} suggested a picture in which BAL quasars represent an early stage in the development of quasars on the basis of their compact radio morphology, observed in most cases.  
\citet{Montenegro-Montes:2008} demonstrated that many radio-emitting BAL quasars share several radio properties common to young radio sources, such as gigahertz-peaked spectrum~(GPS) and compact steep spectrum~(CSS) sources.  
GPS sources are compact ($\lesssim1$~kpc) and show a convex radio spectrum that peaks between 500~MHz and 10~GHz; CSS sources are larger ($\sim1$--20~kpc) and have convex spectra that tend to peak at lower frequencies, typically $<500$~MHz \citep{ODea:1998}.  
On the basis of the rarity of Fanaroff-Riley Class~II radio galaxies in BAL quasars and their observed anticorrelation between the balnicity index and radio loudness, \citet{Gregg:2006} suggested a model in which a BAL phase evolves into a radio-loud phase with a relatively short overlap.  
In this context, AGN wind-induced feedback in the early stages of radio source evolution is discussed for the galaxy--black hole coevolution \citep[e.g.,][]{Lipari:2006,Holt:2008}.    
On the other hand, \citet{Bruni:2012} reported that the fractions of GPS candidates are similar in their BAL and non-BAL quasar samples, suggesting that BAL quasars are generally not younger than non-BAL quasars.  
Thus, neither the orientation scheme nor the evolutionary scheme has been conclusively demonstrated to date on the basis of these inclination indicators and age estimators.

Very-long-baseline interferometry~(VLBI) with milliarcsecond~(mas) angular resolution provides exclusive and crucial opportunities for the investigation of the parsec~(pc)-scale regions in which the phenomena that these inclination indicators and age estimators rely on are actually occuring.     
Several VLBI observations have been reported for BAL-quasar radio sources \citep{Jiang:2003,Kunert-Bajraszewska:2007,Kunert-Bajraszewska:2010,Liu:2008,Doi:2009,Reynolds:2009a,Montenegro-Montes:2009,Gawronski:2010,Yang:2012,Hayashi:2013} and revealed various signatures, such as blazar-like jets with a one-sided morphology, polarized radio emissions, a two-sided morphology suggesting inclined jets, CSS-like characteristics, and interactions between interstellar medium and jet.   
\citet{Doi:2009} reported their systematic VLBI detection survey at 8.4~GHz for 22~radio-loud BAL quasars using the Optically ConnecTed Array for VLBI Exploration project~(OCTAVE: \citealt{Kawaguchi:2008}).  
The samples of bright radio sources ($>100$~mJy~beam$^{-1}$) were selected by position matching between the BAL quasar catalog of \citet{Trump:2006} from the Sloan Digital Sky Survey Third Data Release \citep[SDSS~DR3;][]{Abazajian:2005} and Faint Images of the Radio Sky at Twenty-cm \citep[FIRST;][]{Becker:1995}.    
Most sources~(20/22) were detected with the OCTAVE baselines, suggesting brightness temperatures of greater than 10$^{5}$~K, and the simultaneous coexistence of BAL outflows and nonthermal jets.  Four sources exhibited inverted spectra, suggesting blazars with pole-on-viewed relativistic jets or GPS sources as young radio sources.

In the present paper, we report multifrequency VLBI observations of the BAL quasar J1020+4320 (SDSS J102027.20+432056.2 at $z=1.962$), which was one of the four radio sources showing inverted spectra in the OCTAVE study.  J1020+4320 exhibits BAL troughs of an average velocity of 21341~km~s$^{-1}$ with a width of 1168~km~s$^{-1}$; such a broad width is attributed to an intrinsic absorption.  The absorption index \citep[AI;][]{Hall:2002} is 716~km~s$^{-1}$, according to the modified definition of the AI by \citet{Trump:2006}.  On the other hand, the balnicity index \citep[BI;][]{Weymann:1991} is 0 \citep{Gibson:2009}, according to the criterion of a velocity width $>2000$~km~s$^{-1}$ in the BI definition.  J1020+4320 is a compact radio source in the FIRST image with a resolution of $\sim5\arcsec$.  Previous radio observations in arcsec/arcmin resolutions \citep{Marecki:1999,Vollmer:2008,Orienti:2010a,Stanghellini:2009} suggested that J1020+4320 is a candidate for a high frequency peaker \citep[HFP;][with a spectral peak occurring at frequencies above a few GHz]{Dallacasa:2000}, which is considered to be a younger subclass in the GPS population.  
The present paper is organized as follows.  In Section~\ref{section:observation}, our observations and data reduction procedures are described.  The observational results are presented in Section~\ref{section:result}, and their implications are discussed in Section~\ref{section:discussion}.  
Throughout this paper, a $\Lambda$CDM cosmology with $H_0=70.5$~km~s$^{-1}$~Mpc$^{-1}$, $\Omega_\mathrm{M}=0.27$, and $\Omega_\mathrm{\Lambda}=0.73$ is adopted \citep{Komatsu:2009}.  The comoving distance is 5225~Mpc; 1~milliarcsecond~(mas) corresponds to 8.6~pc at the distance from J1020+4320.

\begin{table*}
\caption{List of Observations\label{table1}}
\begin{center}
\begin{tabular}{clcl} 
\hline\hline
$\nu$ & \multicolumn{1}{c}{Date} & Array & Antenna \\
(GHz) & \multicolumn{1}{c}{} &  & \multicolumn{1}{c}{} \\
(1) & \multicolumn{1}{c}{(2)} & (3) & (4) \\
\hline
1.666 & 2008Mar02 & MERLIN & MK LO CA DE KN DA TA \\
1.658 & 2008Mar02 & EVN & LO Wb Ef On Mc Tr CA Nt \\
2.272 & 2008Aug17 & JVN & VERA$\times$4 Ud \\
6.672 & 2008May04 & JVN & VERA$\times$4 Ud YM \\
8.408 & 2008Aug17 & JVN & VERA$\times$4 Ud Uc Ks \\
\hline
\end{tabular}
\end{center}
\begin{flushleft}
{\footnotesize 
Col.~(1) center frequency: Col.~(2) observation date; Col.~(3) array; Col.~(4) participating antenna.  
MK: Mark~2 $32 \times 25$~m, 
LO: Lovell 76~m, 
CA: Cambridge 32~m, 
DE: Defford 25~m, 
KN: Knocking 25~m, 
DA: Darnhall 25~m, 
TA: Tabley 25~m, 
Wb: Westerbork, Ef: Effelsberg~100~m, On: Onsala~85~m, Mc: Medicina~32~m, Tr: Torun~32~m, Nt: Noto~32~m, VERA: VLBI Exploration of Radio Astrometry~(VERA)~20~m, Ud: JAXA Usuda~64~m, YM: Yamaguchi~32~m, Uc: JAXA Uchinoura~34~m, Ks: Kashima~34~m.}
\end{flushleft}
\end{table*}

\section{OBSERVATIONS AND DATA REDUCTION}\label{section:observation}
VLBI observations at 1.7, 2.3, 6.7, and 8.4~GHz were conducted during a half-year period in 2008~(Table~\ref{table1}).  J1020+4320 was observed at 1.7~GHz using the European VLBI Network~(EVN) and the Multi-Element Radio Linked Interferometer Network~(MERLIN) simultaneously in the snapshot mode.  
The left- and right-circular-polarization signals with a total bandwidth of 32~MHz each for the EVN and 16~MHz each for the MERLIN were obtained; only Stokes-I correlations were used in this study.  Observations at 2.3, 6.7 and 8.4~GHz using the Japanese VLBI Network~(JVN: \citealt{Fujisawa:2008}) were conducted in single circular polarization at a data recording rate of 64~Mbps with 2-bit quantization; this provided an observing band-width of 16~MHz for each band.  The data at 2.3 and 8.4~GHz were obtained simultaneously.  Correlation processing was performed using the Mitaka FX correlator \citep{Shibata:1998} at the National Astronomical Observatory of Japan~(NAOJ).

Data reduction was performed using the Astronomical Image Processing System (AIPS; \citealt{Greisen:2003}).  For the EVN data at 1.7~GHz, amplitude calibration, bandpass calibration, flagging, and fringe-fitting were performed in the standard manner.  The assumed uncertainty of the amplitude calibration is 10\%.  For the JVN data at 2.3, 6.7, and 8.4~GHz, a-priori amplitude calibration was not used because of the lack of a monitoring system of system-noise temperature of several antennas at that time.  The amplitude-gain parameters relative to each antenna were obtained from the self-calibration solution for a point-like strong source, J0958+4725, which was near the target in the sky and was scanned every several tens of minutes to monitor the time variation in the system equivalent flux density~(SEFD) of each antenna.  
The amplitude scaling factor was based on the total flux density of OJ~287, which was measured with the Very Large Array~(VLA) at 8.4~GHz within a week of the JVN observations and Usuda 64-m single-dish observations at 2.3, 6.7, and 8.4~GHz during the JVN observations.  Flux scaling was performed by comparing the self-calibration solutions on the scans of OJ~287\footnote{OJ~287 is one of the rare objects for which almost all of the total flux density can be retrieved at the shortest VLBI baselines according to comparisons between the VLBI correlated flux densities and the total flux densities in the VLA and single-dish monitoring (UMRAO) for many years.} and J0958+4725 at nearly the same elevation; a structure model obtained using the Very Long Baseline Array~(VLBA) was used for the self-calibration of OJ~287.  Although the structure of OJ~287 is known to be variable, it can be assumed to be stable in the range of JVN baselines (less than 50~M$\lambda$ at 8.4~GHz).  
As a check, we found that an amplitude scaling factor obtained from auto-correlation data for methanol masers, whose flux density was determined by a single-dish observation with the Yamaguchi 32~m, showed only a 4\% difference from the OJ~287-based scaling factor \citep{Sugiyama:2011}.  In addition, the flux scaling factor was approximately equal to the factors (with a scattering of less than 10\%) obtained by several other JVN observations \citep{Doi:2006a,Doi:2007,Tsuboi:2008,Sudou:2009,Niinuma:2012}.  We assumed that the uncertainty of the flux densities for our JVN observations is also 10\%.  Imaging was performed using the Difmap software \citep{Shepherd:1994} to apply iteratively deconvolution and self-calibration procedures.

\begin{table*}
\caption{Results of Observations\label{table2}}
\begin{center}
\begin{tabular}{lccccc} 
\hline\hline
$\nu$ & $S_\nu$ & $\sigma$ & $\theta_\mathrm{maj}\times\theta_\mathrm{min}$ & P.A. \\
(GHz) & (mJy) & (mJy beam$^{-1}$) & (mas $\times$ mas) & (deg) \\
(1) & (2) & (3) & (4) & (5)  \\
\hline
  1.666I & $193 \pm 19$  & 0.4  & $136.6\times188.7$ & 48 \\
  1.658I & $164 \pm 16$  & 1.1  & $26.6\times21.3$ & $-87$ \\
  2.272R & $330 \pm 34$  & 3.7  & $16.1\times9.2$ & $-61$ \\
  6.672L & $252 \pm 26$  & 2.7  & $6.0\times3.1$ & $-69$ \\
  8.408R & $206 \pm 21$  & 1.2  & $5.4\times3.3$ & $-52$ \\
 \hline
\end{tabular}
\end{center}
\begin{flushleft}
{\footnotesize
Col.~(1)~center frequency. ``L'' denotes left circular polarization, ``R'' denotes right circular polarization, and ``I'' denotes dual circular polarization; Col.~(2)~flux density; Col.~(3)~image rms noise; Col.~(4)~Half-power beam width; Col.~(5)~position angle of major axis of beam width.
}
\end{flushleft}
\end{table*}

\section{RESULTS}\label{section:result}
\subsection{Radio morphology} 
\begin{figure}
\begin{center}
\FigureFile(1.0\linewidth, ){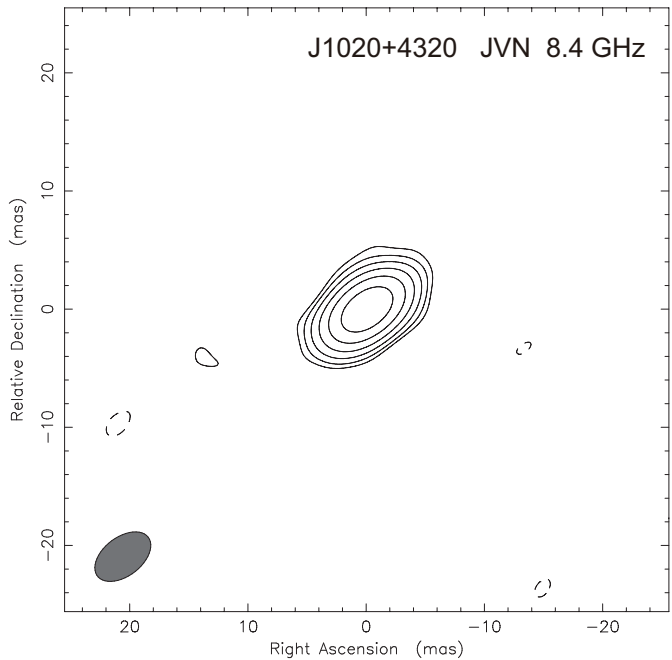}
\end{center}
\caption{JVN image at 8.4~GHz.  Contour levels are $-1$, 1, 2, 4, 8, 16, and 32 times $3\sigma$ of the image rms noise ($\sigma=1.2$~mJy~beam$^{-1}$).}\label{figure1}
\end{figure}

J1020+4320 is nearly unresolved in all our VLBI images at 1.7, 2.3, 6.7, and 8.4~GHz; the measured total flux densities are listed in Table~\ref{table2}.   
At 8.4~GHz (Figure~\ref{figure1}), the deconvolved size determined using the {\tt AIPS} task {\tt JMFIT} in the image domain\footnote{We obtained nearly the same result using the visibility-based {\tt modelfit} in {\tt difmap}.} is $0.6 \pm 0.2$~mas at a position angle~(PA) of $\mathrm{PA} =$~\timeform{87D}~$\pm$~\timeform{55D} and an axis ratio of $\lesssim0.1$, suggesting an elongated structure of $\sim5$~pc.  We also analyzed archival data from the VLBA Imaging and Polarimetry Survey~\citep[VIPS;][]{Helmboldt:2007} at 5~GHz and measured the deconvolved size to be 1.1~mas elongated at $\mathrm{PA} =$~\timeform{67D}, which is consistent with the JVN result.  
The difference in flux density between (simultaneous) MERLIN and EVN observations at 1.7~GHz was $29\pm19$~mJy, which was minor ($\sim15$\%) as compared to the total emission (193~mJy).  Thus, most of the emission is concentrated in a central region within $\sim 25$~mas (the beam size of EVN) corresponding to $\sim200$~pc.  An unresolved structure in the MERLIN image constrains the entire size to $<190$~mas, corresponding to $<1.6$~kpc.

\subsection{Radio spectrum} 
\begin{figure}
\begin{center}
\FigureFile(1.0\linewidth, ){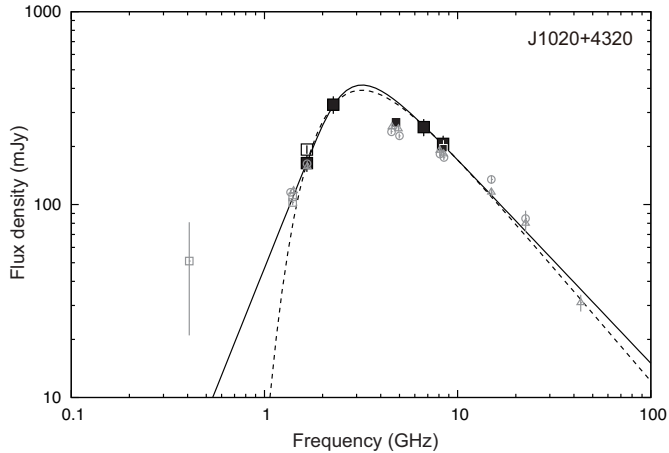}
\end{center}
\caption{Radio spectrum of J1020+4320.  
Large filled squares and a large open square: our VLBI and MERLIN measurements, respectively, in 2008.  
Small filled squares: OCTAVE at 8.4~GHz on November~04, 2007 \citep{Doi:2009} and VIPS at 5~GHz on May~01, 2006 \citep{Helmboldt:2007}.  
Small open circles: quasi-simultaneous VLA on June 25, 1999 \citep{Stanghellini:2009}.  
Small open triangles: quasi-simultaneous VLA in January, 2004 \citep{Orienti:2010a}.  
Small open squares: NVSS on November~15, 1993 \citep{Condon:1998} and FIRST on February~20, 1997 \citep{Becker:1995} at 1.4~GHz and MERLIN at 408~MHz from November 1994 to January 1995 \citep{Marecki:1999}.  
Solid and dashed curves indicate spectra fitted to our four-frequency VLBI data (large filled squares) with spectral models of synchrotron self-absorption and free--free absorption, respectively (Section~\ref{section:result}).
}
\label{figure2}
\end{figure}

Our VLBI results suggest a spectrum with a turnover (Figure~\ref{figure2}).  
We applied spectral fitting to {\it only} our 2008 VLBI data at four frequencies (see Tables~\ref{table1} and \ref{table2}).  
A simple power-law spectral model including synchrotron self-absorption~(SSA), $S_\nu = S_0 \nu^{2.5} [1-\exp{(-\tau \nu^{\alpha_0 - 2.5})]}$, provided a peak flux density $S_\mathrm{p} = 417$~mJy at a peak frequency $\nu_\mathrm{p}=3.2$~GHz and $\alpha_0=-1.1$.  
A spectral model including free--free absorption~(FFA), $S_\nu = S_0 \nu^\alpha_0 \exp{(-\tau \nu^{-2.1})}$, provided $S_\mathrm{p} = 392$~mJy at $\nu_\mathrm{p}=3.2$~GHz\footnote{The spectral peak frequencies of J1020+4320 have been determined to be $\sim1$--5~GHz in a series of previous studies as well \citep{Marecki:1999,Vollmer:2008,Stanghellini:2009,Orienti:2010a}, which are consistent with our fitting results.  These minor discrepancies are attributed to their different model functions, (such as a broken power-law, hyperbola, and parabola) which do not have any direct physical significance.} and $\alpha_0=-1.2$.    
Although these spectral fittings were unable to discriminate conclusively between SSA and FFA, peak frequencies of $\nu_\mathrm{p}=3.2$~GHz are suggested in either case.

\subsection{Variability} 
Our VLBI measurements and the two fitted spectra are in good agreement with previous studies (small symbols in Figure~\ref{figure2}), even those with different beam sizes.   
Therefore, J1020+4320 is a compact and stable radio source without significant variability over the last decade.

\section{DISCUSSION}\label{section:discussion}
The observational results confirmed a series of characteristics corresponding to a young radio source: (1)~compact morphology, (2)~a giga-hertz peaked spectrum, and (3)~little variability \citep[][for a review]{ODea:1998}.  
Therefore, we conclude that the BAL quasar J1020+4320 possesses a young radio source at its nucleus.

The determined spectral peak frequency of 3.2~GHz in the observer frame is equivalent to a peak at 9.5~GHz in the rest frame at $z=1.962$.  
The age estimator based on the observed correlation between the linear size and peak frequency of CSS and GPS radio sources \citep{ODea:1997,Snellen:2000} indicates that the radio source in J1020+4320 might be extremely young with an age of the order of $\sim100$~years ($\sim10$~pc in size at a tentative expansion velocity of $\sim0.3c$; see \citealt{ODea:1997}).  
Indeed, the marginally deconvolved structure with an elongation of $\sim1$~mas in the JVN 8.4-GHz and VIPS 5-GHz images (Section~\ref{section:result}) is suggestive of a mini radio galaxy of $\sim10$~pc across.  
Alternatively, the result of SSA spectral fitting suggests a source diameter of the order of $\sim10$~pc (with a magnetic field of $\sim0.1$~G) 
under the near equipartition condition between the energy densities of radiating electrons and magnetic field in a homogeneous, self-absorbed, incoherent synchrotron radio source with a power-law electron energy distribution.  
This linear size is also consistent with our VLBI images.  
Thus, we can naturally understand the observed radio properties of J1020+4320 in terms of its compactness (with a size of $\sim10$~pc).  
To confirm that the radio source is actually young, further investigations might be essential for the determination of its kinematic age by, e.g., measuring expected spectral changes on spectral property undergoing adiabatic expansion \citep{Orienti:2010a}.

On the other hand, the data point at 408~MHz ($51 \pm 30$~mJy with MERLIN; \citealt{Marecki:1999}) is significantly distant from our fitted model spectra.  One more component is necessary to fit the low-frequency emission.  
It may originate in an optically thin component at large scales with a steep spectrum, which can also be inferred from the observed $29\pm19$~mJy at 1.7~GHz (the MERLIN--EVN differential flux density) from the region of $\sim0.2$--1.6~kpc (Section~\ref{section:result}).  
The low-frequency component could be conceivably be explained as a relic of past jet activity \citep[][and references therein]{Baum:1990,Saikia:2009}.  
In the case of J1020+4320, the age of the GPS component is $\sim100$~years, while the low-frequency component may be relic of jets emanated $\sim0.7$--$5 \times 10^4$~years ago (assuming an expansion rate of $0.1c$).  
The coexistence of components with distinct ages indicates recurrent jet activity.  
The current stage of 3C~84 may be the closest example for the restarted jet activity in the recent 100~years in an existing radio galaxy \citep{Asada:2006,Nagai:2009,Nagai:2010}; the classical double-lobed radio galaxy 1245+676 also contains a GPS component of an expanding compact double with a separation of 14~pc, suggesting a kinematic age of $\sim200$~years \citep{Marecki:2003}.    
Similarly, J1020+4320 may not be so young as a radio source but contains a very young radio component originating in the restarted jet activity.

Composite spectra with GPS and MHz components have been found in a proportion of radio-loud BAL quasar \citep{Montenegro-Montes:2008,Bruni:2012} as well as J1020+4320.     
The fact that the restarted jets are associated with BAL features suggests relevance to the evolutionary scheme for the origin of BAL quasars \citep{Hayashi:2013}.  
In the evolutionary scheme, the BAL fraction in the quasar population \citep[$\sim$20\%; e.g.,][]{Hewett:2003,Knigge:2008} requests the duration of the BAL phase of $\sim2 \times 10^7$~years in the AGN lifetime of $\sim10^8$~years.         
On the other hand, time scales of $\sim10^6$~years for the intermission of the jet have been inferred in several double-double radio galaxies such as PKS~B1545$-$321 \citep{Safouris:2008}, 4C~02.27 \citep{Jamrozy:2009}, B1834+620 \citep{Schoenmakers:2000}, and Cygnus~A \citep{Steenbrugge:2008}, although various different time scales have also been inferred from existing observations \citep[][for a review]{Saikia:2009}.   
Thus, because the time scale between successive episodes of the jet activity are possibly much shorter than the BAL phase, short-lived GPS sources may appear repeatedly during the BAL phase.     
J1020+4320 may be one of the quasars with a recently reactivated jet during a BAL phase.  
At this stage, our study on a quasar in the BAL--GPS composite phase constitutes the first step toward additional extensive research rather than the conclusive evidence for the relationship between the processes initiating/interrupting BAL outflow and a nonthermal jet for the galaxy--black hole coevolution.

\bigskip
We thank the anonymous referee for the constructive suggestions that have improved the clarity of the paper.  We are grateful to all the staffs and students involved in the development and operation of the Japanese VLBI network~(JVN).  The JVN project is led by the National Astronomical Observatory of Japan~(NAOJ), which is a branch of the National Institutes of Natural Sciences~(NINS), Yamaguchi University, Hokkaido University, Gifu University, Kagoshima University, Tsukuba University, Osaka Prefecture University, and Ibaraki University, in collaboration with the Geographical Survey Institute~(GSI), the Japan Aerospace Exploration Agency~(JAXA), and the National Institute of Information and Communications Technology~(NICT).  The observations on August~17, 2008, in this study was the first astronomical VLBI observation for the JAXA Uchinoura 34~m antenna.  MERLIN is operated by the University of Manchester as a National Facility of the Science and Technology Facilities Council (STFC).  The European VLBI Network is a joint facility of European, Chinese and other radio astronomy institutes funded by their national research councils.  We used the US National Aeronautics and Space Administration's (NASA) Astrophysics Data System~(ADS) abstract service and NASA/IPAC Extragalactic Database (NED), which is operated by the Jet Propulsion Laboratory~(JPL).  In addition, we used the Astronomical Image Processing System~(AIPS) software developed at the National Radio Astronomy Observatory~(NRAO), a facility of the National Science Foundation operated under cooperative agreement by Associated Universities, Inc.  This research has made use of data from the University of Michigan Radio Astronomy Observatory which has been supported by the University of Michigan and by a series of grants from the National Science Foundation, most recently AST-0607523.  This study was partially supported by Grants-in-Aid for Scientific Research (C; 21540250 and B; 24340042, AD) from the Japan Society for the Promotion of Science (JSPS).

\end{document}